# Noise Models in the LISA Mission

Michele Pagone and Carlo Novara

**Abstract**

This document briefly describes the noise models and shapes used for the synthesis of the Drag-Free and Attitude Control System in the LISA space mission. LISA (Laser Interferometer Space Antenna) is one of the next large-class missions from the European Space Agency (ESA), expected to be launched in 2034. The main goal of the mission is to detect the gravitational waves, which are undulatory perturbations of the space-time fabric, extremely important to collect experimental proofs for the General Relativity Theory. In the 90s, different international collaborations of institutes laid the foundations for the first ground-based interferometers (see, e.g., LIGO and Virgo). However, ground-based interferometers have a limited bandwidth due to the Earth's environmental noises and short arm-length of few kilometers. Therefore, they cannot observe gravitational waves belonging to the portion of the spectrum below 1 Hz. This issue can be overcome by means of space-based interferometers, that can have arm-lengths up to millions of kilometers and exploit a quieter environment than the Earth's surface. The LISA system is affected by actuation, sensing and environmental disturbances and noises. Among the actuation noises we have those given by the Micro Propulsion System (MPS), the Gravitational Reference Sensor (GRS) and the Optical Assembly (OA) motor. Among the sensing noises we consider the interferometer, the Differential Wavefront Sensor (DWS) and the GRS. The environmental disturbances are given by the solar radiation pressure, the test-mass stiffness and self-gravity, and the environmental noises acting directly on the test-mass.

## Actuation and Sensing Noises

A generic noise can be modelled as follows:

$$d = H(s)w$$

where $H(s)$ is a suitable zero-pole filter that approximates the noise spectral density function of the actuator/sensor and $w$ is a white noise. Therefore, the GRS forces and torques noises ($\boldsymbol{d}_E$ and $\boldsymbol{D}_E$, respectively) are:

$$\boldsymbol{d}_E = \begin{bmatrix} H_{HRa_x}(s)w_x \\ H_{HRa_{yz}}(s)w_y \\ H_{HRa_{yz}}(s)w_z \end{bmatrix} \quad \boldsymbol{D}_E = \begin{bmatrix} H_{HRa_t}(s)w_x \\ H_{HRa_t}(s)w_y \\ H_{HRa_t}(s)w_z \end{bmatrix}$$

where $H_{HRa_x}, H_{HRa_{yz}}, H_{HRa_t}$ are the noise shape filters which approximate the spectral density functions of the GRS actuation noise:

$$H_{HRa_{yz}} = 5 \cdot 10^{-15} \frac{(s + 1.257 \cdot 10^{-4})^2}{(s + 2.81 \cdot 10^{-6})^2} \frac{\text{N}}{\sqrt{\text{Hz}}}$$

$$H_{HRa_t} = 5 \cdot 10^{-17} \frac{(s + 1.257 \cdot 10^{-4})^2}{(s + 2.81 \cdot 10^{-6})^2} \frac{\text{Nm}}{\sqrt{\text{Hz}}}$$

Michele Pagone and Carlo Novara are with the Department of Electronics and Telecommunications, Politecnico di Torino, Corso Duca degli Abruzzi, 24, 10129 Torino, Italy. {michele.pagone, carlo.novara}@polito.it.

In the specific case of the drag-free mode, the longitudinal electrodes are off, therefore there is no noise along the x-coordinate. So, the $\boldsymbol{d_E}$ term becomes:

$$\boldsymbol{d_E} = \begin{bmatrix} 0 \\ H_{HRa_{yz}}(s)w_y \\ H_{HRa_{yz}}(s)w_z \end{bmatrix}$$

The corresponding noises are plotted in Figure 1 and Figure 2.

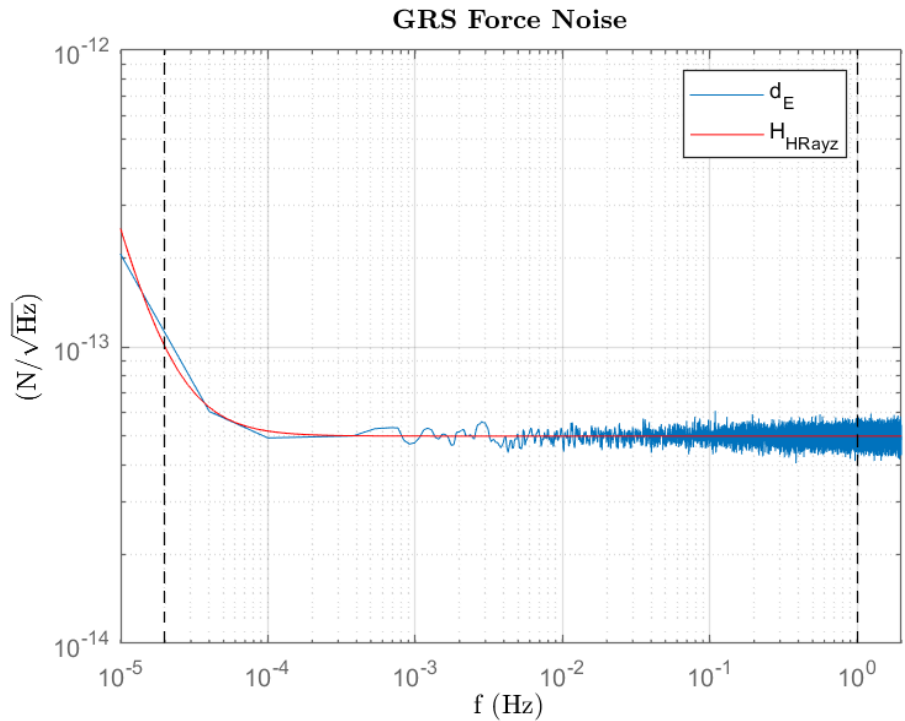

**Figure 1 – GRS Force Noise (y-z component). Blue: simulated, red: shape filter.**

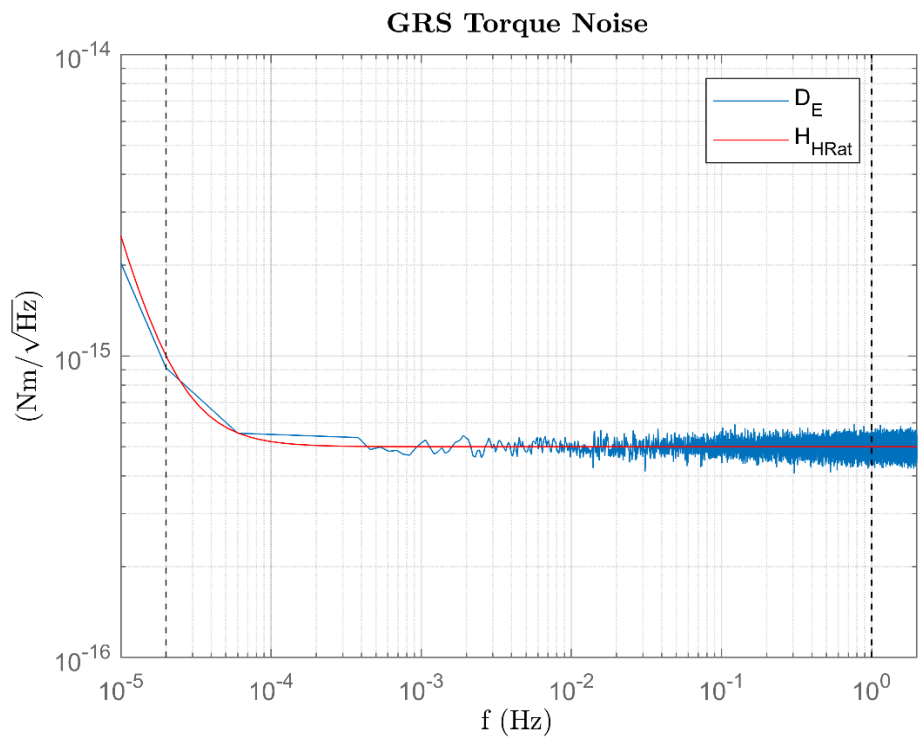

**Figure 2 – GRS Torque Noise. Blue: simulated, red: shape filter.**

For what concerns the thrusters' noise, the spectral density function can be approximated by:

$$H_T = 10^{-7} \frac{(s + 6.283 \cdot 10^{-2})^2}{(s + 8.886 \cdot 10^{-3})^2}.$$

Single thruster noises are described by the following relations and depicted in Figures 3-4.

$$H_{MPS_{Fx}} = H_{MPS_{Fy}} = 1.3\, H_T \qquad H_{H_{MPS_{Fz}}} = 2.2\, H_T$$

$$H_{H_{MPS_{Mx}}} = H_{H_{MPS_{My}}} = 2.3\, H_T \qquad H_{H_{MPS_{Mz}}} = 1.4\, H_T$$

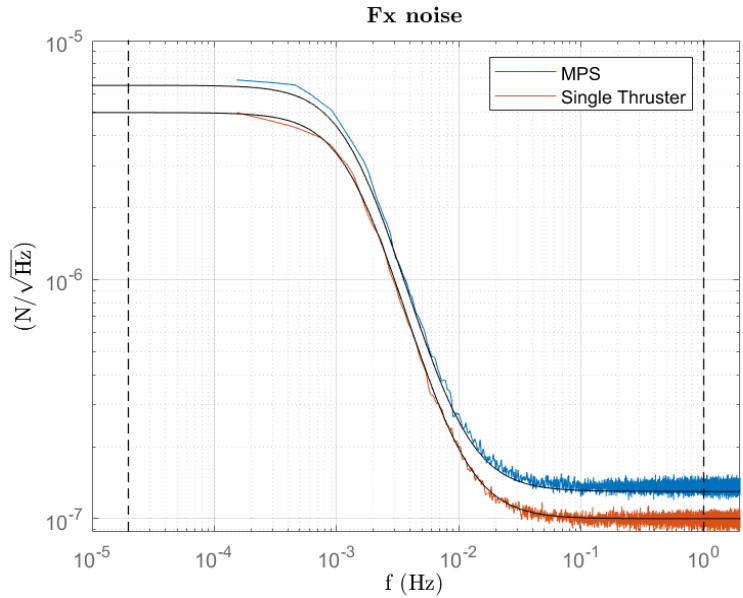

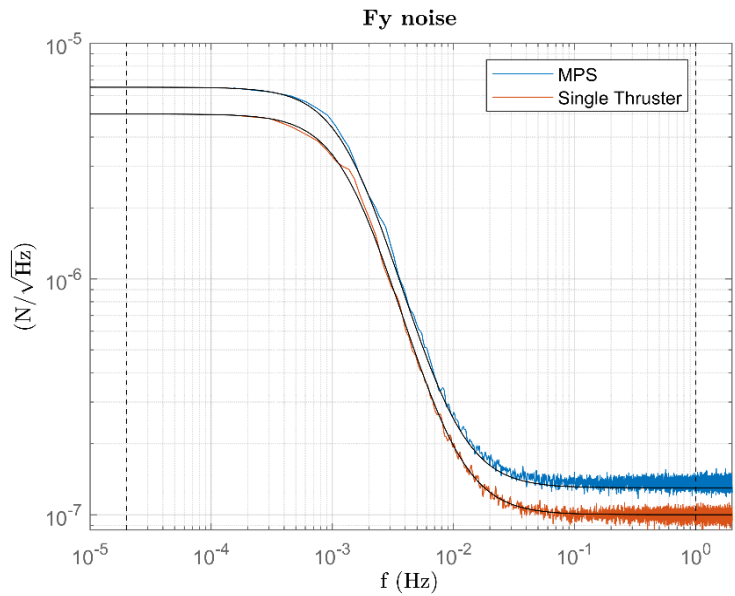

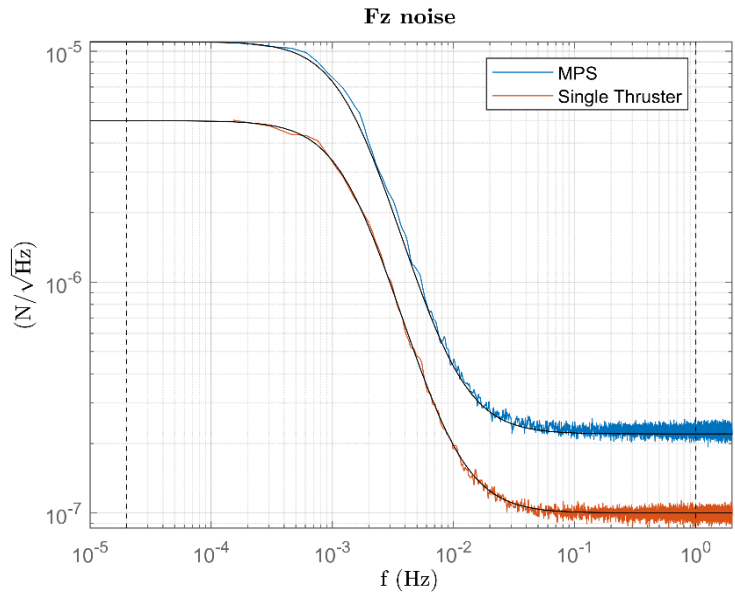

**Figure** Errore. Nel documento non esiste testo dello stile specificato. **– MPS force noises. Red: single thruster. Blue: MPS system**

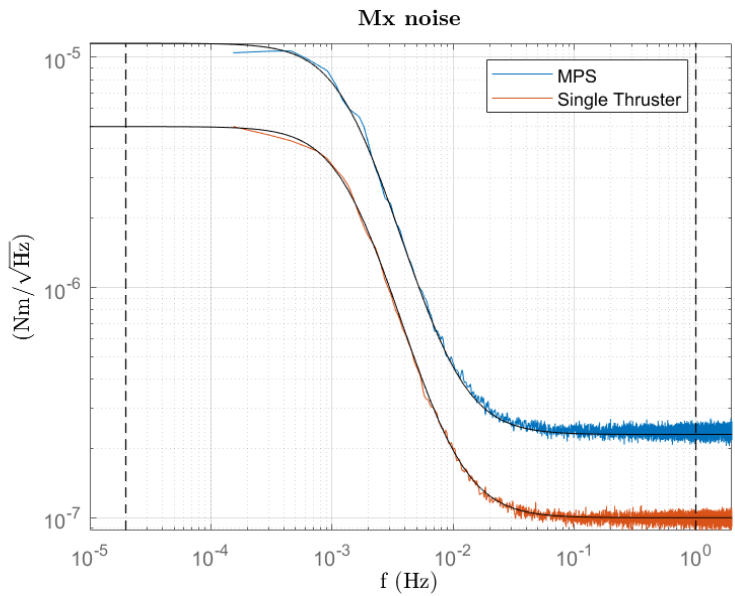

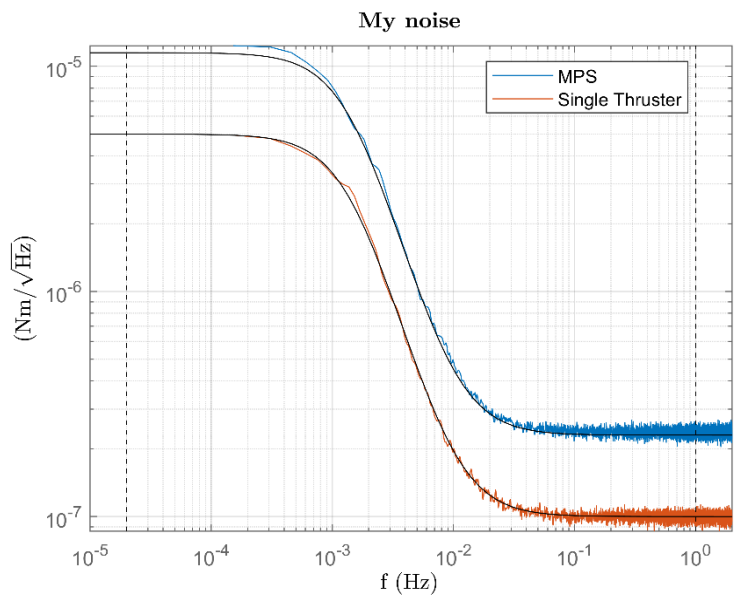

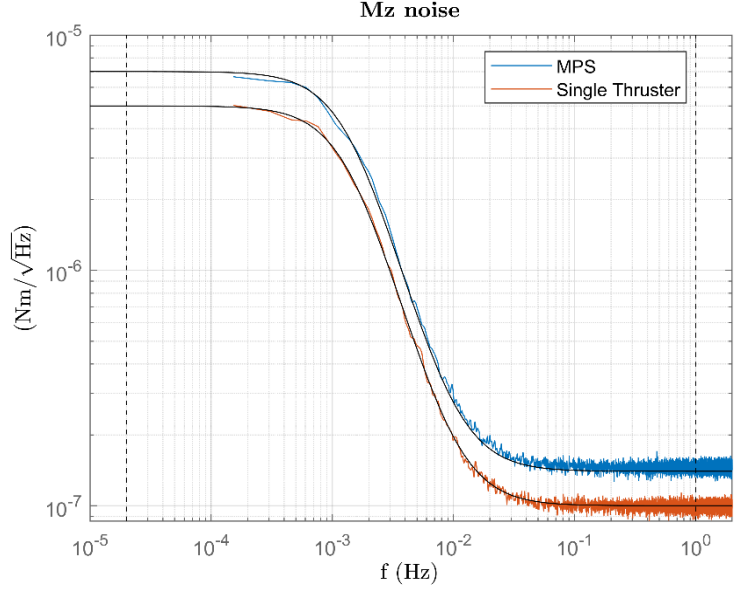

**Figure 4 – MPS torque noises. Red: single thruster. Blue: MPS system**

For what concerns the sensing noises in the drag-free mode, the longitudinal position of the test mass relative to the cage center is measured by the local interferometer, while the lateral and vertical position by the GRS. Hence, the measured test-mass position is given by:

$$\tilde{r}_M = r_M + d_{r_M} = r_M + \begin{bmatrix} H_{IFO} w_x \\ H_{HRs_{xy}} w_y \\ H_{HRs_z} w_z \end{bmatrix}$$

where

$$H_{IFO} = 1.5 \cdot 10^{-12} \frac{(s + 1.3 \cdot 10^{-2})^2}{(s + 1 \cdot 10^{-4})^2} \frac{m}{\sqrt{Hz}}$$

$$H_{HRs_{xy}} == 1.8 \cdot 10^{-9} \frac{(s + 3 \cdot 10^{-2})(s + 5.4 \cdot 10^{-3})(s + 9.6 \cdot 10^{-4})(s + 1.7 \cdot 10^{-4})}{(s + 2.58 \cdot 10^{-2})(s + 2.933 \cdot 10^{-3})(s + 4.333 \cdot 10^{-4})(s + 6 \cdot 10^{-5})} \frac{m}{\sqrt{Hz}}$$

$$H_{HRs_z} == 3 \cdot 10^{-9} \frac{(s + 3 \cdot 10^{-2})(s + 5.4 \cdot 10^{-3})(s + 9.6 \cdot 10^{-4})(s + 1.7 \cdot 10^{-4})}{(s + 2.58 \cdot 10^{-2})(s + 2.933 \cdot 10^{-3})(s + 4.333 \cdot 10^{-4})(s + 6 \cdot 10^{-5})} \frac{m}{\sqrt{Hz}}$$

are the noise shapes of the interferometer and of the GRS position sensor, depicted in In Figure 5, Figure 6 and Figure 7.

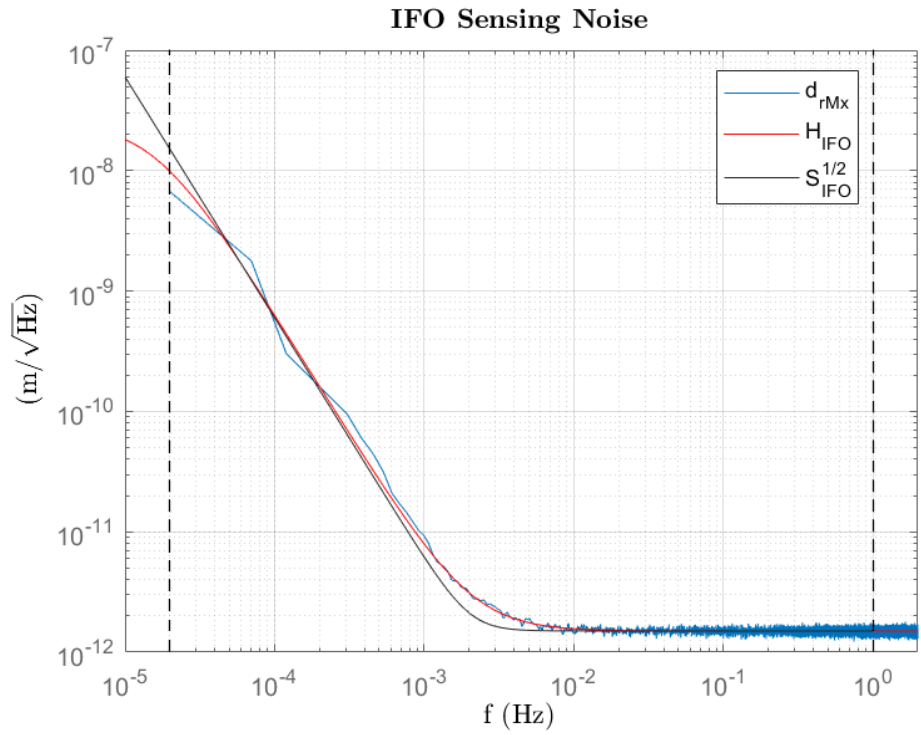

**Figure 5 – Interferometer sensing noise. Blue: simulated, red: shape filter**

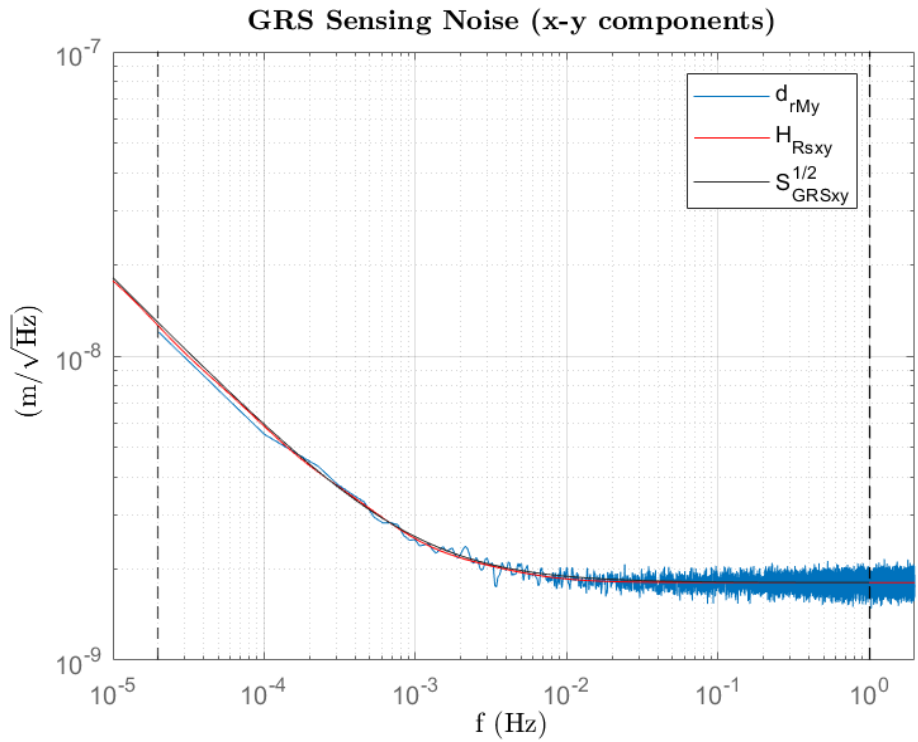

**Figure 6 – GRS Sensing Noise (longitudinal and lateral position). Blue: simulated, red: shape filter**

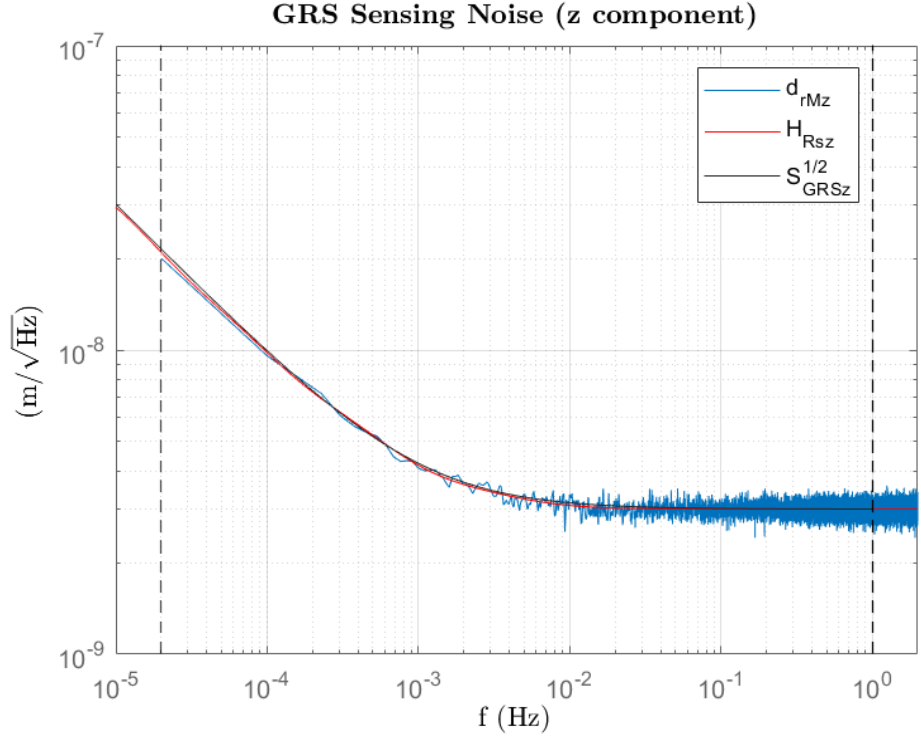

**Figure 7 - GRS sensing noise (vertical position). Blue: simulated, red: shape filter**

In drag-free mode, the test mass attitude is measured by the DWS on the pitch and yaw coordinates. The roll coordinate can only be measured by the GRS. Hence, the measured TM attitude is

$$\tilde{\boldsymbol{\vartheta}}_M = \boldsymbol{\vartheta}_M + \boldsymbol{d}_{\vartheta_M} = \boldsymbol{\vartheta}_M + \begin{bmatrix} H_{HRstx} w_x \\ H_{DWS} w_y \\ H_{DWS} w_z \end{bmatrix}$$

where $H_{HRstx}$ is the GRS roll noise shape filter and $H_{DWS}$ is DWS shape filter:

$$H_{HRs_{tx}} = 2 \cdot 10^{-7} \frac{(s + 3 \cdot 10^{-2})(s + 5.4 \cdot 10^{-3})(s + 9.6 \cdot 10^{-4})(s + 1.7 \cdot 10^{-4})}{(s + 2.58 \cdot 10^{-2})(s + 2.933 \cdot 10^{-3})(s + 4.333 \cdot 10^{-4})(s + 6 \cdot 10^{-5})} \frac{\text{rad}}{\sqrt{\text{Hz}}}$$

$$H_{DWS} = 5 \cdot 10^{-9} \frac{(s + 6 \cdot 10^{-3})^2}{(s + 1 \cdot 10^{-5})^2} \frac{\text{rad}}{\sqrt{\text{Hz}}}.$$

We conclude with the DWS-SC noise, which corrupts the azimuth-elevation angles of the spacecraft relative to the incoming laser beams:

$$\tilde{\varphi} = \varphi + H_{DWS\_SC} w_\varphi \qquad \tilde{\psi} = \psi + H_{DWS\_SC} w_\psi$$

$$H_{DWS\_SC} = 1.167 \cdot 10^{-10} \frac{(s + 6 \cdot 10^{-3})^2}{(s + 6 \cdot 10^{-5})^2} \frac{\text{rad}}{\sqrt{\text{Hz}}}.$$

Such noises are plotted in Figure 8, Figure 9, and Figure 10.

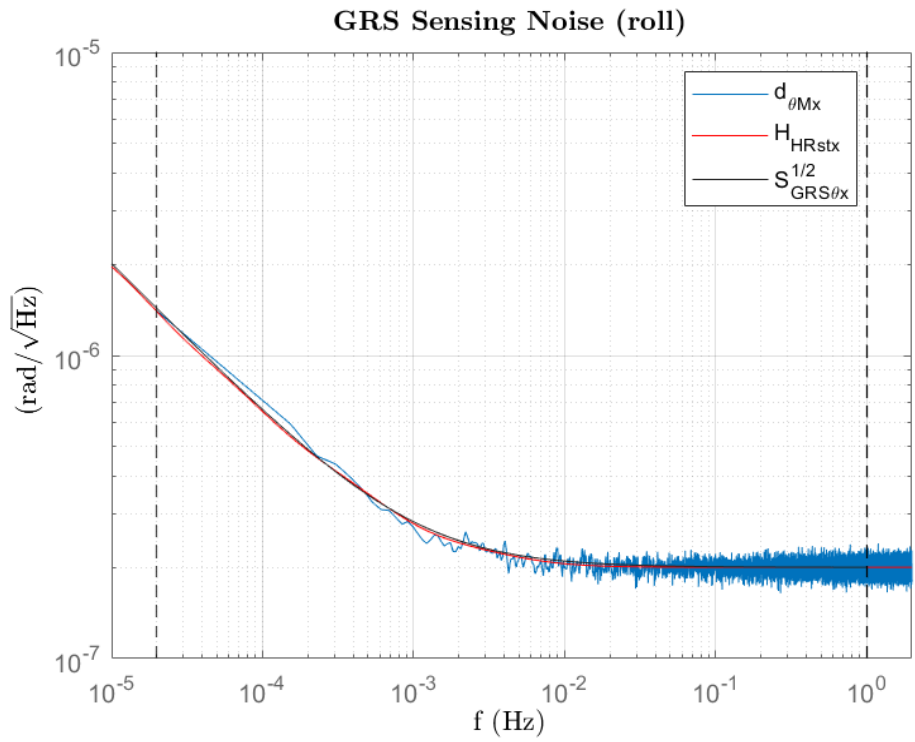

**Figure 8 – GRS sensing noise (test-mass roll). Blue: simulated, red: shape filter**

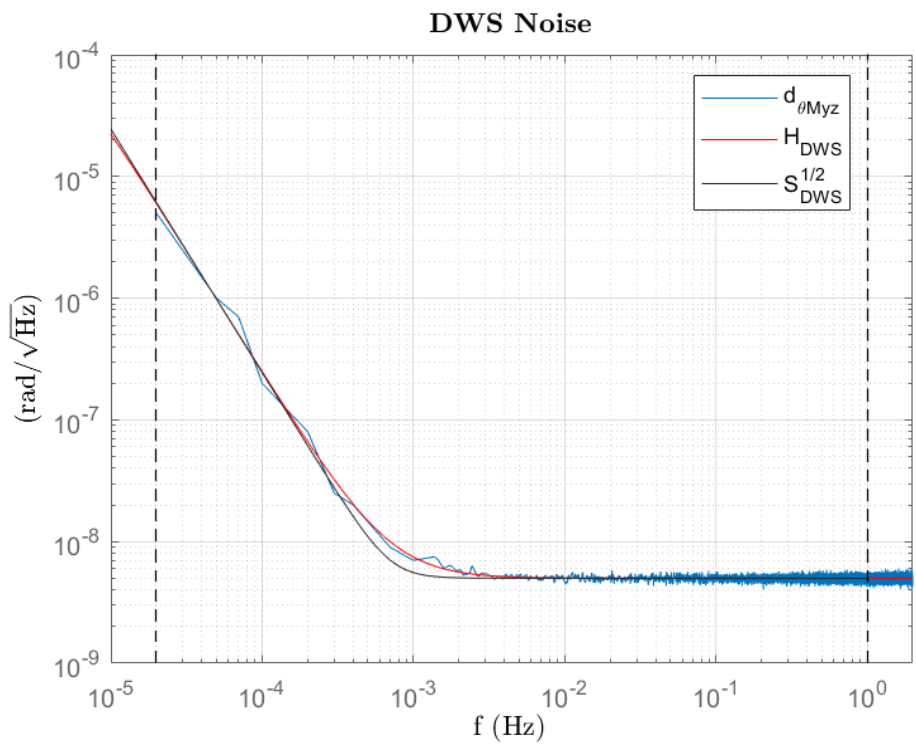

**Figure 9 – DWS sensing noise (test-mass pitch/yaw). Blue: simulated, red: shape filter**

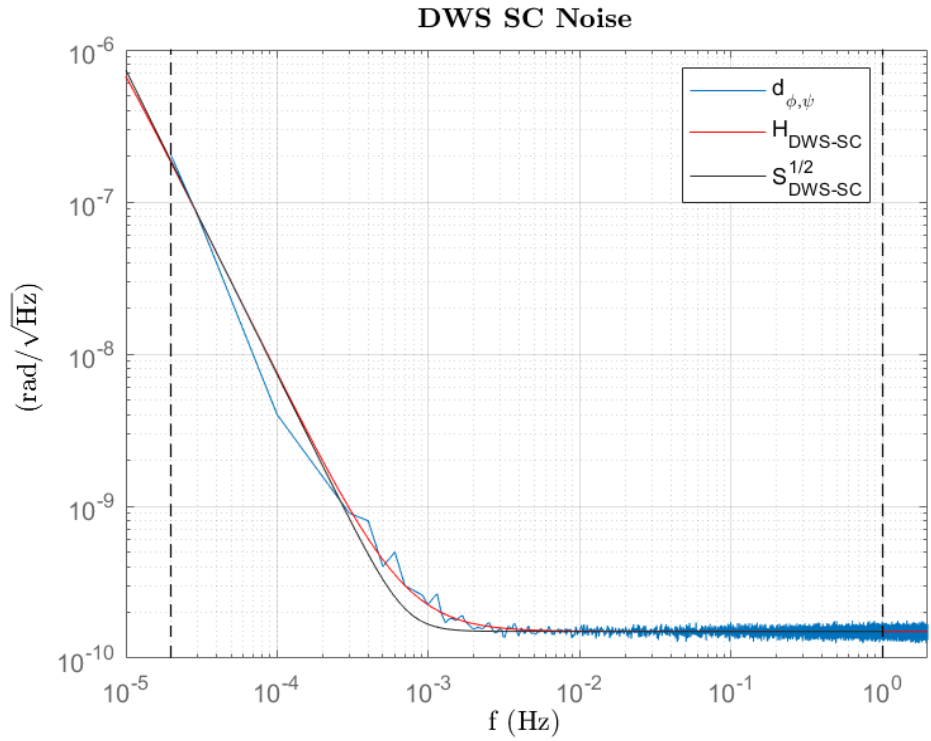

**Figure 10 – DWS SC sensing noise (azimuth-elevation). Blue: simulated, red: shape filter**

## Solar Radiation Pressure

One of the disturbances acting on the LISA spacecraft is the solar radiation pressure (see Figure 11). The solar flux is affected by the noise $d_{sn}$:

$$d_{sn} = H_{SP} w$$

$$H_{SP} = 7.87 \cdot 10^{-11} \frac{(s + 7.09 \cdot 10^{-2})(s^2 + 5.78 \cdot 10^{-3} s + 2.954 \cdot 10^{-4})}{(s + 4.712 \cdot 10^{-3})(s^2 + 4 \cdot 10^{-3} s + 4 \cdot 10^{-4})} \frac{N}{\sqrt{Hz}}$$

where w is a white noise.

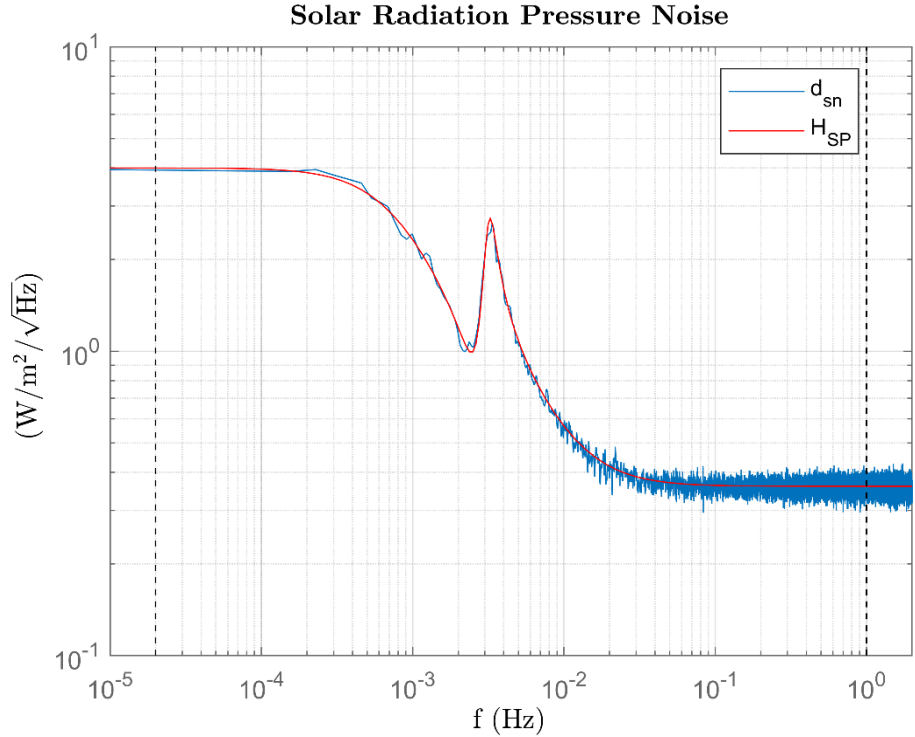

**Figure 11 – Solar flux noise. Blue: simulated, red: shape filter**

## TM Environmental Noises

The direct disturbances acting on the test-mass are given by a set of contributions that can be collected into a single spectral density function. Hence, the environmental noise acting on a test mass is given by:

$$\boldsymbol{d}_{env} = \begin{bmatrix} H_{TMd} w_x \\ H_{TMd} w_y \\ H_{TMd} w_z \end{bmatrix} \quad \boldsymbol{D}_{env} = \begin{bmatrix} H_{TMD} w_x \\ H_{TMD} w_y \\ H_{TMD} w_z \end{bmatrix}$$

where $H_{TMd}$ is a transfer function that approximates the spectral density, while $H_{TMD} = l_m H_{TMd}$ is the noise shape for the disturbance torque where $l_m = 0.46$ m is the test-mass side length. The transfer functions $H_{TMd}$ and $H_{TMD}$ are:

$$H_{TMd} = 1.07 \cdot 10^{-15} \frac{(s + 9 \cdot 10^{-3})(s + 1.62 \cdot 10^{-3})(s + 2.88 \cdot 10^{-4})(s + 5.1 \cdot 10^{-5})}{(s + 7.74 \cdot 10^{-3})(s + 8.88 \cdot 10^{-4})(s + 1.3 \cdot 10^{-4})(s + 1.8 \cdot 10^{-5})} \frac{N}{\sqrt{Hz}}$$

$$H_{TMD} = 4.92 \cdot 10^{-17} \frac{(s + 9 \cdot 10^{-3})(s + 1.62 \cdot 10^{-3})(s + 2.88 \cdot 10^{-4})(s + 5.1 \cdot 10^{-5})}{(s + 7.74 \cdot 10^{-3})(s + 8.88 \cdot 10^{-4})(s + 1.3 \cdot 10^{-4})(s + 1.8 \cdot 10^{-5})} \frac{Nm}{\sqrt{Hz}}$$

Such spectral densities are plotted in Figure 12 and Figure 13.

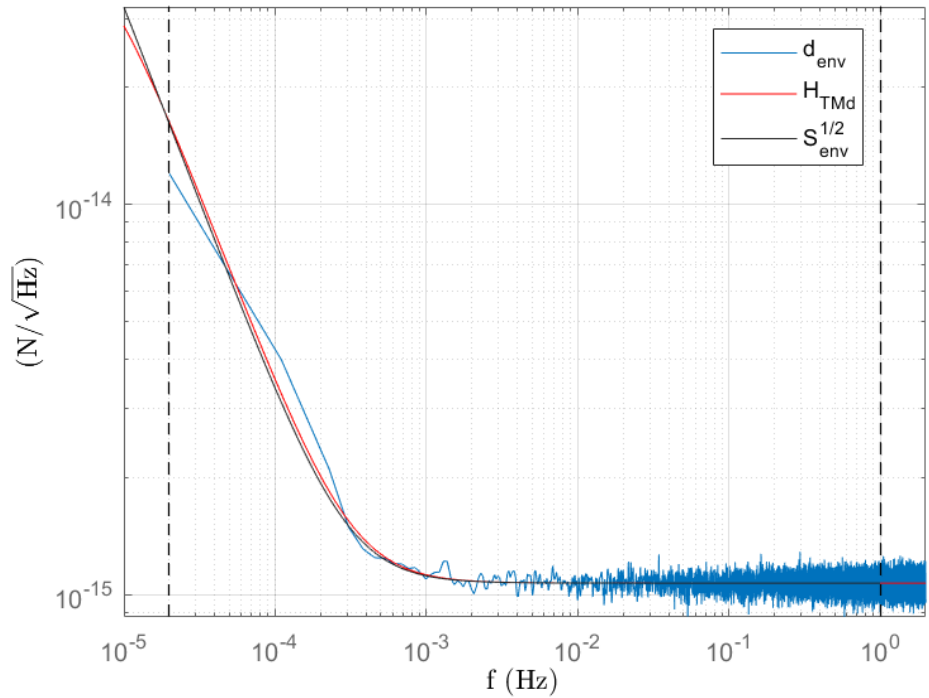

**Figure 12 – Enviromental force noise acting on test-mass (single component)**

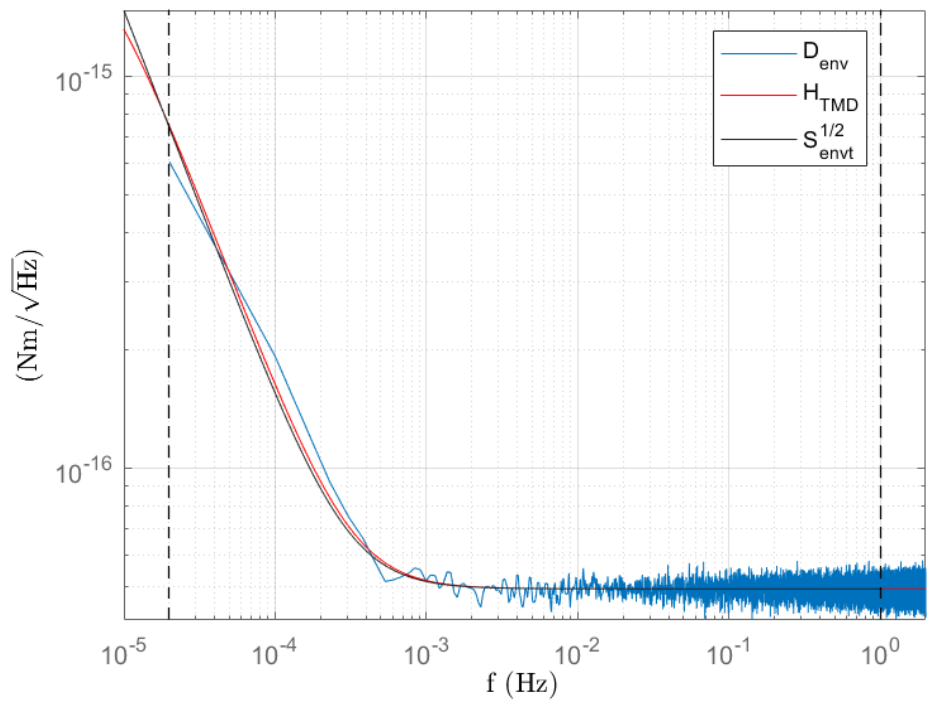

**Figure 13 – Enviromental torque noise acting on test-mass (single component)**